\documentclass[useAMS,usenatbib]{mn2e}

\usepackage{graphicx}
\usepackage{amssymb}
\usepackage[hyphens]{url}

\newcommand{\apj}{ApJ}

\newcommand{\aap}{A\&A}

\newcommand{\mnras}{MNRAS}

\newcommand{\MC}{\multicolumn}
\newcommand{\kms}{km~s$^{-1}$}

\newcommand{\HI}{H{\sc i}}
\newcommand{\HII}{H{\sc ii}}
\newcommand{\sunn}{$_{\odot}$}
\newcommand{\acc}{atoms~cm$^{-2}$}
\newcommand{\logOH}{12+\log(\textrm{O/H})}

\title[Study of the isolated galaxy UGC~4722]
{Study of the Lynx-Cancer void galaxies. - V. The extremely isolated
galaxy UGC~4722}
\author[J.~Chengalur, S.~Pustilnik, D.~Makarov,
Y.~Perepelitsyna, E.~Safonova, I.~Karachentsev ]
{J.N.~Chengalur,$^{1}$\thanks{E-mail: chengalur@ncra.tifr.res.in
(JNC)} S.A.~Pustilnik,$^{2}$\thanks{E-mail: sap@sao.ru (SAP)}
D.I.~Makarov,$^{2}$
Y.A.~Perepelitsyna,$^{2}$ \newauthor
E.S.~Safonova,$^{3}$
I.D.~Karachentsev$^{2}$ \\
$^1$ National Center for Radio Astrophysics, TIFR, Pune, India \\
$^2$ Special Astrophysical Observatory of RAS, Nizhnij Arkhyz,
  Karachai-Circassia 369167, Russia\\
$^3$ Sternberg Astronomical Institute, Lomonosov Moscow State University,
  13 Universitetsky pr., Moscow, 119991, Russia
}

\begin{document}

\label{firstpage}

\date{Accepted on 2014 October ??. Received on 2014 October 2}

\pagerange{\pageref{firstpage}--\pageref{lastpage}} \pubyear{2014}

\maketitle

\begin{abstract} We present a detailed study of the extremely isolated 
Sdm galaxy UGC~4722 ($M_{\rm B} = -17.4$)  located in the nearby Lynx-Cancer 
void. UGC~4722 is a member of the catalogue of isolated galaxies, and has
also been identified as one of the most isolated galaxies in the Local
Supercluster. Optical images of the galaxy however show that it has
a peculiar morphology with an elongated  $\sim$14~kpc-long plume.
New observations with the Russian 6-m telescope (BTA)
and the Giant Metrewave Radio Telescope (GMRT) of the ionised and neutral gas
in UGC~4722 reveal the second component responsible for the disturbed
morphology of the system. This is a small, almost completely destroyed,
very gas-rich dwarf ($M_{\rm B} = -15.2$, $M$(\HI)/$L_{\rm B} \sim4.3$)
We estimate the oxygen abundance for both galaxies to be $\logOH\sim7.5-7.6$
which is 2--3 times lower than what is expected from the 
luminosity-metallicity relation for similar galaxies in denser environments.
The $ugr$ colours of the plume derived from Sloan Digital Sky Survey
(SDSS) images are consistent with a simple stellar population with
a post starburst age of 0.45--0.5~Gyr. This system hence appears to be 
the first known case of a minor merger with a prominent tidal
feature consisting of a young stellar population.
\end{abstract}


\section[]{INTRODUCTION}
\label{sec:intro}

Detailed studies of void galaxies allow one to determine the effect of
the environment on the evolution of galaxies. The isolation and (assumed)
low rate of interactions and mergers make void galaxies a useful population
for testing models of galaxy formation and evolution. The possibility that 
galaxies in voids have significantly different properties than galaxies in
denser regions has attracted considerable attention \citep[e.g.,][and 
references therein]{Peebles01, Gottlober03, Hoeft06, Hoeft10, Hahn07, 
Hahn09, Kreckel2011a}.  However, most previous studies have been
focused on  large distant ($d \gtrsim$ 100-200~Mpc) voids
\citep[][among others]{Rojas05, Patiri06, Sorrentino06},
drawn from the SDSS and 2dFRGS surveys, and consequently are biased towards
more luminous galaxies. These studies generally found at best moderate  
differences between galaxies in voids and walls. Numerical
simulations \cite[e.g.][]{Kreckel2011a} predict that  luminous dwarfs
(M$_{\rm r}$  $\lesssim -18$) in voids are statistically
indistinguishable from similar dwarfs in higher density regions. However, 
fainter dwarfs (M$_{\rm r}$  $\gtrsim -16$) are expected to be significantly 
bluer and have
to have higher specific star formation rates than their higher density 
counterparts. Since, as mentioned above, previous studies have included 
very few of these faint systems, there is a clear need to extend the 
observational samples to include low-mass galaxies.

In order to study the effect of the environment on the properties of faint
dwarfs, one would have to focus on nearby regions, where current facilities
have sufficient sensitivity to detect faint dwarfs. In the surroundings
of the Local Volume (i.e. the region within 10~Mpc from the Local Group)
one can probe the properties of galaxies ten to hundred times less luminous
than the galaxies studied in the distant voids. Motivated by this consideration
\citet{PaperI} drew up a sample of galaxies in the nearby Lynx-Cancer void.
The gas phase metallicity and photometrical properties of the void galaxy
population  have been presented in \citet{void_OH,void_photo}.
Detailed studies of some peculiar galaxies located in this region
can be found in \citet{HS0822,DDO68,void_LSBD,CP2013}. The galaxies studied
in the last cited works are unusual in that they have a very large
$M$(\HI)/$L_{\rm B}$ ratio, unusually blue colours, and/or very low
metallicity. Other galaxies found in nearby voids include LSBDs
J0015+0104
\citep{Eridanus} and J0926+3343 \citep{J0926} which have the 2$^{nd}$ and 
3$^{rd}$  lowest metallicities known (after SBS~0335--052~W).

In this paper we present a detailed study of the Lynx-Cancer void galaxy 
UGC~4722.  It is included in the Catalog of Isolated Galaxies (CIG) 
\citep{CIG} as CIG~293 
(KIG~293 in the nomenclature of the NASA/IPAC Extragalactic Database). The 
catalog was built by total visual inspection of the Palomar Sky Survey films
for galaxies brighter than 15.7 mag. \citet{CIG} proposed the simple but 
effective criterion of isolation using only the angular diameters of galaxies.
The isolation criteria requires the absence of `significant' neighbours around 
the considered galaxy.  We have also checked the SDSS image the region of 
$\pm25^{\prime}$ (roughly $\pm200$ kpc in projection) surrounding the galaxy, 
but do not 
find any object that could be a possible companion to this system. Finally, 
UGC~4722 is one of 520 galaxies identified as being the most isolated, 
`orphan' galaxies found in the Local Supercluster \citep{ikar11}
among galaxies with known redshift. Despite being isolated
UGC~4722 appears morphologically disturbed and has a long tail extending 
from the northern side of the disk (see Fig.~\ref{fig:image}).  \citet{ikar11} 
listed it as one of 21 `orphan' galaxies with a peculiar structure. UGC~4722 
shows clear signs of interaction, but without a visible source for the 
disturbance.  \citet{ikar06,ikar08} found 8 such spatially isolated galaxies
with signs of distortion,  and based on the suggestions for the existence
of ``dark galaxies'' by \citet{vandenberg1969} and \citet{trentham2001},
proposed that these could be cases where a galaxy is interacting with 
a companion consisting entirely of dark matter. 

The rest of this paper is organised as follows. In Sec.~2 we describe the 
observations of this galaxy and the data reduction. They include \HI\
21-cm line observations with  the GMRT (Giant Meterwave Radio Telescope) and 
long-slit spectra obtained using the 6-meter telescope (BTA) of the Special 
Astrophysical Observatory of the Russian Academy of Sciences (SAO RAS).
SDSS (Sloan Digital Sky Survey) imaging data and their reduction are 
presented as well.
In Sec.~3 we present the results of GMRT and BTA observations, as well as
the result of photometry of the SDSS images. In Sec.~4 we discuss the
observational
properties of this unusual galaxy and discuss interpretations of the
observations.

\section[]{OBSERVATIONS AND DATA REDUCTION}

\label{sec:obs}

\subsection{Optical observations}

\begin{table*} [hbtp]
\begin{center}
\caption{Journal of the 6\,m telescope observations of UGC~4722 system}
\label{tab:journal}
\begin{tabular}{lrccccccrr} \\ \hline \hline
\MC{1}{c}{ Date }       &
\MC{1}{c}{ Expos. }   &
\MC{1}{c}{ Wavelength [\AA] } &
\MC{1}{c}{ Dispersion } &
\MC{1}{c}{ Spec.resol. } &
\MC{1}{c}{ Seeing }     &
\MC{1}{c}{ Airmass }     &
\MC{1}{c}{ Grism }       &
\MC{1}{c}{ Slit}         &
\MC{1}{c}{ PA }       \\

\MC{1}{c}{ }       &
\MC{1}{c}{ time [s] }    &
\MC{1}{c}{           } &
\MC{1}{c}{ [\AA/pixel] } &
\MC{1}{c}{ FWHM(\AA) } &
\MC{1}{c}{ [arcsec] }    &
\MC{1}{c}{          }    &
\MC{1}{c}{          }    &
\MC{1}{c}{ No.}          &
\MC{1}{c}{ ($\degr$)}    \\

\MC{1}{c}{ (1) } &
\MC{1}{c}{ (2) } &
\MC{1}{c}{ (3) } &
\MC{1}{c}{ (4) } &
\MC{1}{c}{ (5) } &
\MC{1}{c}{ (6) } &
\MC{1}{c}{ (7) } &
\MC{1}{c}{ (8) } &
\MC{1}{c}{ (9) } &
\MC{1}{c}{ (10) } \\
\hline
\\[-0.3cm]
 2006.11.18  & 2$\times$600 & $ 5700-7400$ & 0.9 & 5.5  & 1.5 & 1.11 & VPHG1200R & 2 & 13  \\
 2006.11.18  & 2$\times$600 & $ 5700-7400$ & 0.9 & 5.5  & 1.5 & 1.08 & VPHG1200R & 3 & 30  \\
 2012.11.17  & 1$\times$1200& $ 5700-7400$ & 0.9 & 5.5  & 1.7 & 1.06 & VPHG1200R & 1 & 4  \\
 2013.01.15  & 3$\times$900 & $ 3500-7300$ & 2.0 & 12.0 & 2.0 & 1.06 & VPHG550G  & 1 & 4  \\
\hline \hline \\[-0.2cm]
\end{tabular}
\end{center}
\end{table*}

\begin{figure}
 \centering
 \includegraphics[angle=-0,width=7.5cm]{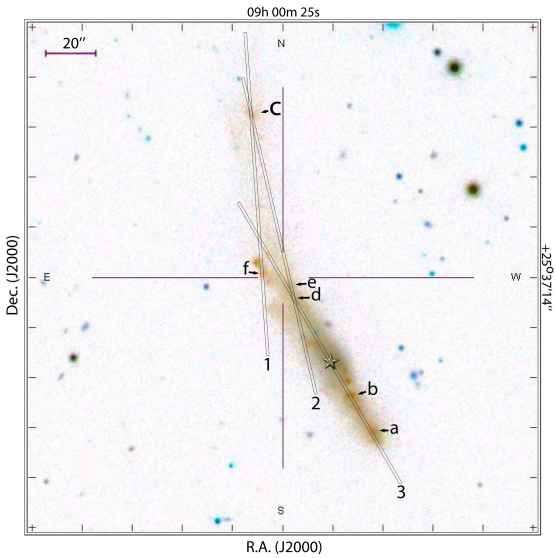}
  \caption{\label{fig:image}
The SDSS $gri$ image (inverse colours, size $\sim$200\arcsec) of the system
UGC~4722/UGC~4722C, with the BTA long slits shown superimposed for the
observations on 18/Nov/2006 and 15/Jan/2013 (No.~1, PA = 4\degr)
and 18/Nov/2006 (No.~2, PA = 13\degr\ and No.~3, PA = 30\degr). Arrows 
with letters indicate prominent features seen in the P--V diagrams (Fig.~2 
and 3). The star shows the approximate position of the galaxy centre.
 This corresponds to the location of the steep velocity gradient in the
P--V diagram in Fig.~3 (bottom panel).
}
\end{figure}

Two long-slit spectra of UGC~4722 were obtained using the BTA and the 
multimode focal reducer SCORPIO in combination with grisms
VPHG1200R and VPHG550G and a 2K$\times$2K CCD detector EEV-42-40.
The main observational parameters are given in Tab.~\ref{tab:journal}.
The observations made with the VPHG1200R grism are used for kinematical
studies while the observations made with the other grim are used for
abundance
determinations. The slit width was 1\arcsec, and the slit length was 
$\sim 6$\arcmin. Pixels along the slit were binned to give an effective 
pixel size of 0.36\arcsec. The slit position (which was used for observations
with both grisms) is shown in Fig.~\ref{fig:image} superposed on the SDSS 
image of the UGC~4722 region (with a field size of  $\sim$200\arcsec).
As can be seen from Fig.~\ref{fig:image} the slit (labelled ``1'') was aligned 
with the diffuse plume, and also covered the compact emission-line nebulosity
{\bf `C'} at the northern edge of the plume as well as the  bright \HII-region
{\bf `f'} at the northern end of the disk. 

Archival BTA long slit observations were also available for this galaxy. 
The observations were taken on 18/Nov/2006 using grism VPHG1200R. 
The rest of the instrumental set up was identical to that described above.
Observations were obtained for two slit positions, shown in 
Fig.~\ref{fig:image} ({labelled as ``2'' and ``3''}).  As can be 
seen from
Fig.~\ref{fig:image}, slit position 3 with a position angle of 
30\degr\ is aligned with the major axis of the disk, and intersects 
the bright \HII-regions {\bf a,b}. Slit position 2 with a position 
angle of 13\degr\ goes along the edge of the plume. It includes
the nebulosity  {\bf `C'} at the edge of the plume and also intersects the
northern edge of the disk in between the two bright \HII-regions {\bf d,e}.

The spectra were processed using standard procedures, as described, e.g. in
\citet{DDO68}. Position-Velocity (P-V) diagrams for the H$\alpha$ emission line
were derived using a {\tt MIDAS}\footnote{MIDAS is an acronym for the 
European Southern -- Munich Image Data Analysis System. } program for 
reduction of long-slit spectra written by Makarov and Kniazev. The details 
can be found in \citet{Zasov00}. For abundance determination from the 
spectrum with grism VPHG550G the intensities of all important emission 
lines were measured using standard procedures in the MIDAS package 
\citep{Kniazev08}.

\subsection{HI observations}

GMRT HI 21~cm observations of UGC~4722 were carried out on 18/Nov/2013.
The correlator was configured to a total bandwidth of 4.17~MHz
($\sim$890~\kms)
divided into 512 channels (or a velocity resolution of 1.74~\kms) centred
at the heliocentric redshift of the galaxy. The initial flagging and 
calibration  were carried out using the FLAGCAL pipeline \citep{flagcal,
flagcal1}, and the subsequent processing was done using the AIPS package. 
A continuum image was made using the line free channels and used for 
self-calibration. The self calibration solutions were then applied to the
line visibilities, and the continuum emission subtracted out using the 
task UVSUB. Images were then made at a variety of resolutions using the 
task IMAGR, and residual continuum subtracted out using the task IMLIN. 
Images of the integrated \HI\ emission and the \HI\ velocity field were 
made using the task MOMNT.

\begin{table}
\caption{Parameters of the GMRT observations}
\label{tab:obspar}
\begin{tabular}{ll}
\hline
     & UGC~4722  \\
\hline
Date of observations     & 2013 Nov 18  \\
Field center R.A.(2000)  &09$^{h}$00$^{m}$23.54$^{s}$   \\
Field center Dec.(2000)  &+25$^{o}$36$^{'}$40.6$^{"}$    \\
Central Velocity (\kms)  & 1750.0   \\
Time on-source  (h)      &$\sim$6  \\
Number of channels       & 256  \\
Channel separation (\kms)& $\sim$1.73 \\
Flux Calibrators         & 3C48,3C286 \\
Phase Calibrators        & 0842+185 \\
Resolution (arcsec$^{2}$) (rms (mJy~Bm$^{-1}$)) & 41~$\times$~38 (2.2)\\
                                              & 26~$\times$~23 (1.9)\\
                                              & 12~$\times$~11 (1.6)\\
\hline
\hline
\end{tabular}
\end {table}

\subsection{Photometry from the SDSS images}


We made photometric measurements using the public archive of the
SDSS Data Release 7 \citep[DR7;][]{DR7}. Since the SDSS provides users with
fully reduced images, the only additional step we needed to perform
was background subtraction. For this, all bright stars were removed from 
the images. After that the target object was masked and the background 
level was estimated using the {\tt aip} package from {\tt MIDAS}. The 
photometry of galaxies was then performed in circular apertures. A more 
detailed description of this method and the related programs can be found 
in \citet{Kniazev04}. To transform instrumental fluxes in apertures to 
stellar magnitudes, we used the photometric system coefficients defined 
in SDSS for the relevant fields. The accuracy of the zero-point determination 
was  $\sim$0.01 mag in all filters.

\section[]{RESULTS}
\label{sec:results}

\subsection[]{Long-slit observations}

\begin{figure*}
  \centering
 \includegraphics[angle=-0,width=15.0cm, clip=]{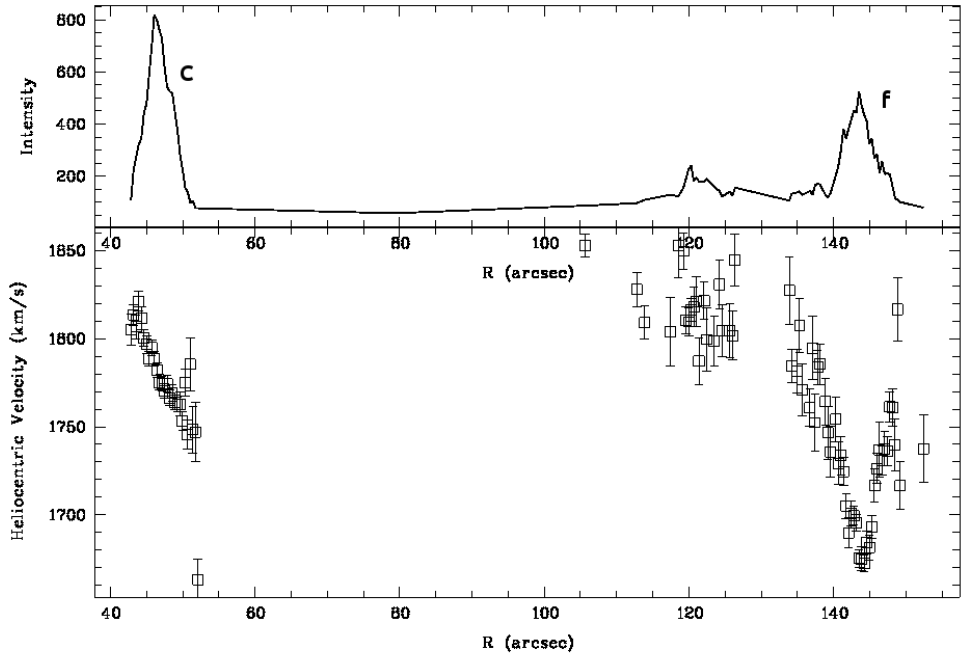}
  \caption{
Slit position No.~1.
Top panel: Distribution of H$\alpha$-line intensity along the slit (north 
           is in the left) in arbitrary units;
Bottom panel: Position-Velocity (P--V) diagram. For both panels the X-axis 
              shows the position along the slit in arcsec. The Y-axis shows 
              the relative intensities and radial velocity in \kms 
              respectively. Letters `C' and `f' on H$\alpha$ intensity peaks
              denote to corresponding objects in Fig.~\ref{fig:image}
}
	\label{fig:PV}
 \end{figure*}

\begin{figure*}
  \centering
\includegraphics[angle=-0,width=15.0cm, clip=]{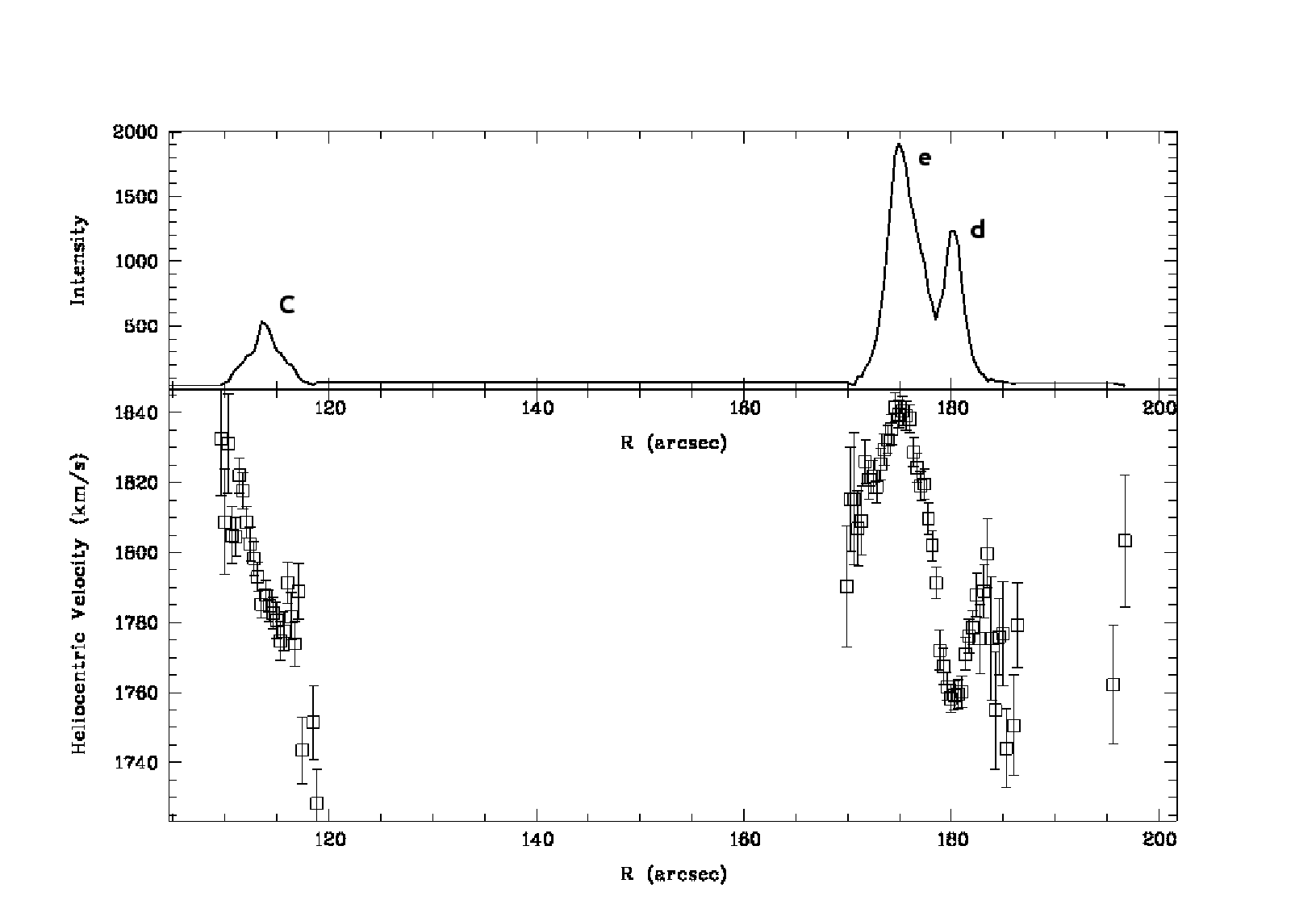}
\includegraphics[angle=-0,width=15.0cm, clip=]{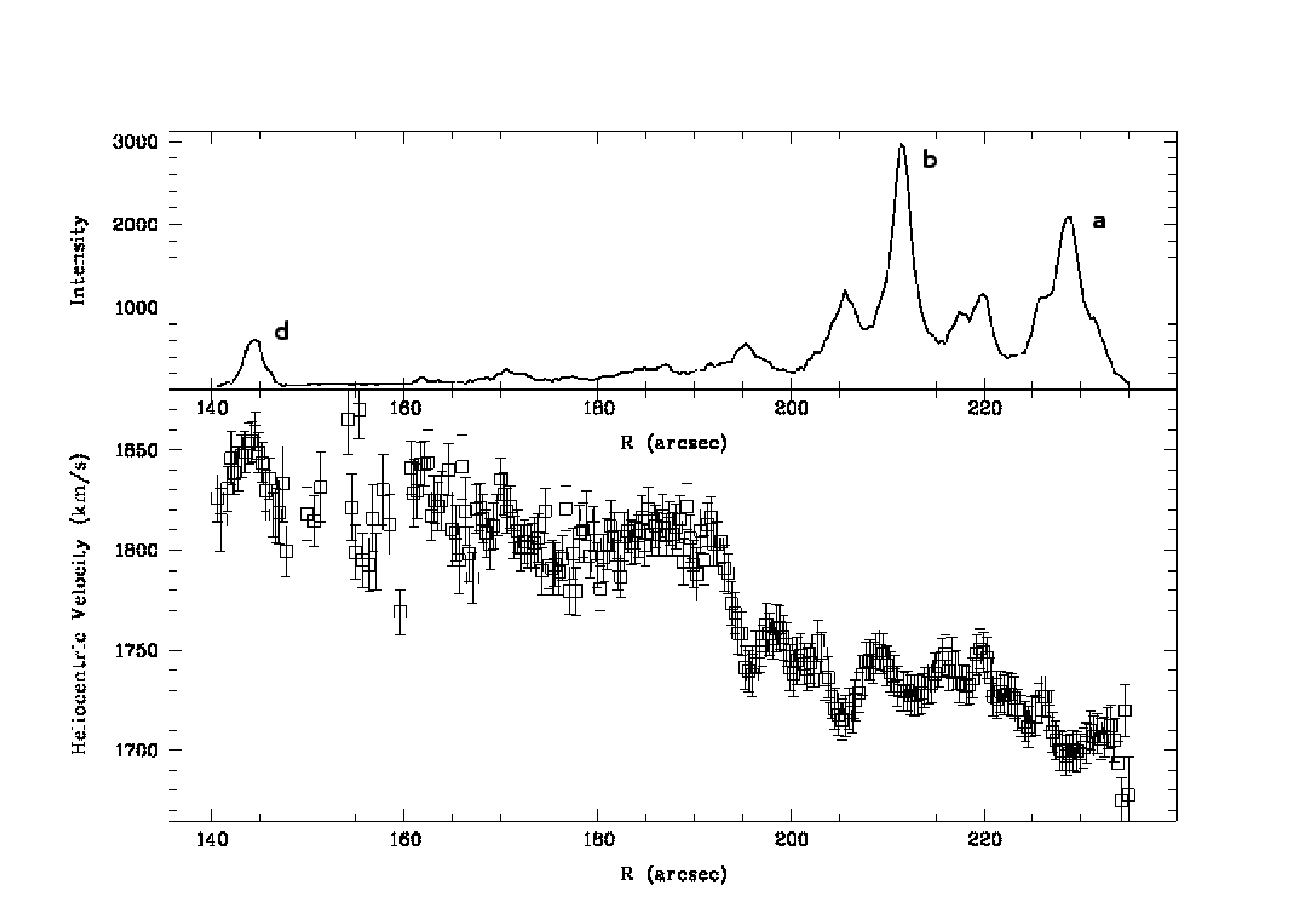}
  \caption{Same as in Fig.~\ref{fig:PV}, but for observations on 18/Nov/2006.
Top panel: Slit position No.~2. The distribution of H$\alpha$ intensity 
           (arbitrary units) and the P--V diagram. The compact nebulosity `C'
           is centered at $X \sim$114\arcsec.
Bottom panel: Slit position No.~3. Letters `C', `a' - `e' on the H$\alpha$ 
            intensity peaks refer to the corresponding objects in 
            Fig.~\ref{fig:image}. 
}
	\label{fig:PV2006}
 \end{figure*}

In Fig.~\ref{fig:PV} we show the H$\alpha$-line intensity distribution (in 
relative units, solid line ) along the long slit position No.~1 (top panel) 
and the 
corresponding radial velocity in \kms\ (bottom panel) for regions where the
signal-to-noise ratio is greater than 3.0 and the error in the estimated
velocity $\sigma_{\rm V}$ is less than  20~\kms. The intensity is computed
within a narrow wavelength range 
(width $\delta \lambda$ = 10~\AA) centred at the redshifted H$\alpha$-line.
Letters `C' and 'f' near the H$\alpha$ intensity peaks refer to the 
corresponding objects in Fig.~\ref{fig:image}. The total distance along the 
slit is about 110\arcsec. The left edge of the plot corresponds to the 
northern edge of the slit which probes the ionised gas motions in 
nebulosity~`C' at the plume edge. The size of this nebulosity is 
$\sim 8$\arcsec, and the slit direction is inclined at 
${\mathrm PA}_{\mathrm 1}\sim$30\degr\ with respect to the major axis of 
the nebulosity. The right edge of the plot corresponds to the NE part of the 
main galaxy disc with the slit direction of ${\mathrm PA}_{\mathrm 2} \sim$25\degr\ with 
respect to the galaxy major axis. 

As can be seen from the bottom panel of the figure, the velocity of the 
ionised gas in component `C' is centred at $V_{\rm hel}$(C) $\sim$1780~\kms, 
and has a spread of $\sim$75~\kms. The variation of velocity along the
slit resembles a solid body rotation curve.  For the main galaxy the 
velocity curve is disturbed, with a distinct velocity swing close to 
the position of the most prominent \HII\ region (marked here as
region `f'). The form and the sign of this swing is consistent with it arising
from an expanding ionised shell around the off-plane \HII\ region with maximal
velocities of $\sim$60--80~\kms. Similar cases of expanding H$\alpha$ shells
were presented for e.g. in \citet{VV2}; physical models for these shells are
discussed by \citet{WB99}. The ionised gas outside of the shell
has a velocity of $\sim$1810~\kms, with a spread of $\pm$20--30~\kms.

In Fig.~\ref{fig:PV2006} we present similar data for observations from the
BTA run on 18/Nov/2006. In the top panel we show the distribution of the H$\alpha$
intensity and radial velocity for slit position No.~2 (along the northern part
of the main galaxy disc, plume and component `C'). The positions of the
prominent \HII-regions `C', `d' and `f' are also marked. Similar velocity 
swings with amplitudes of 30--40~\kms\ are seen near the latter two regions.
For component `C' the `solid-body' like rotation is seen with full amplitude
range of $\sim$60~\kms. In the bottom panel similar data are shown
for slit position No.~3 (i.e. along the main galaxy major axis). There are
several prominent \HII-regions in the southern part of the disc (including
`a' and `b'), and smaller regions in the northern part, including `d' on the
edge. The velocity curve shows a steep gradient near the centre of the galaxy
(i.e. abcissa between 190 and 195), which flattens as one approaches the 
edge of the visible H$\alpha$ extent of the galaxy (i.e. $\sim$95\arcsec). 
As in the previous plots, multiple velocity swings are seen near positions 
of prominent \HII-regions, indicative of expanding shells. Accounting for
these swings, the total rotation velocity amplitude is $\sim$130~\kms\,
somewhat smaller than the velocity spread seen in the \HI\ (see Sec.~\ref{ssec:hi} below). We also note 
that the H$\alpha$ velocities for the central part of UGC~4722 appear to 
suffer from a systematic zero-point shift of $\sim$20--30~\kms with 
respect to the \HI\ velocities described below. This systematic error is 
within the expected internal velocity calibration error for the optical 
long slit spectra. 

\subsection[]{\HI\ distribution and velocity field}
\label{ssec:hi}

\begin{figure*}
  \centering
\includegraphics[angle=0,width=7.5cm, clip=]{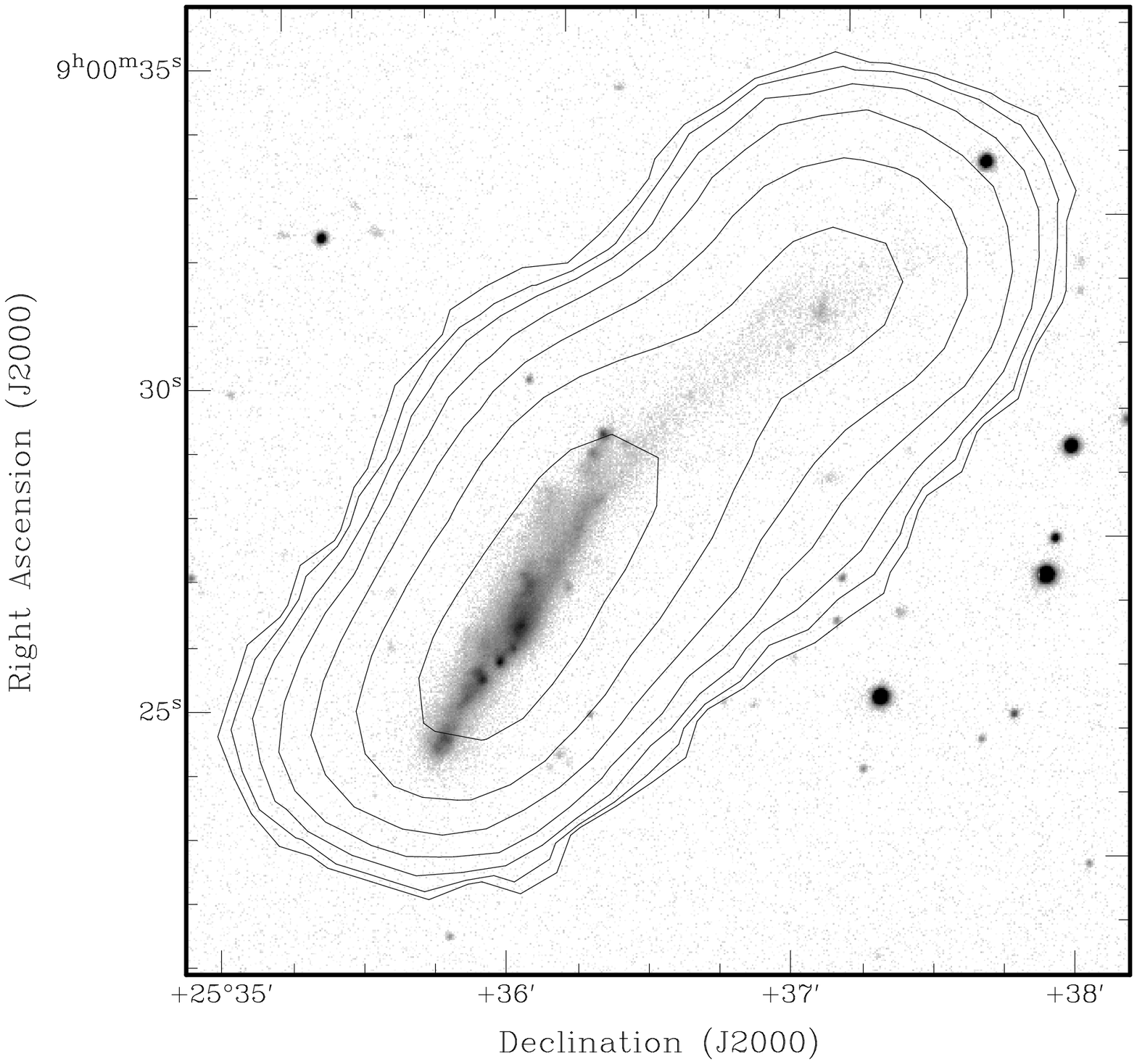} \hskip 0.5in \includegraphics[angle=0,width=7.5cm, clip=]{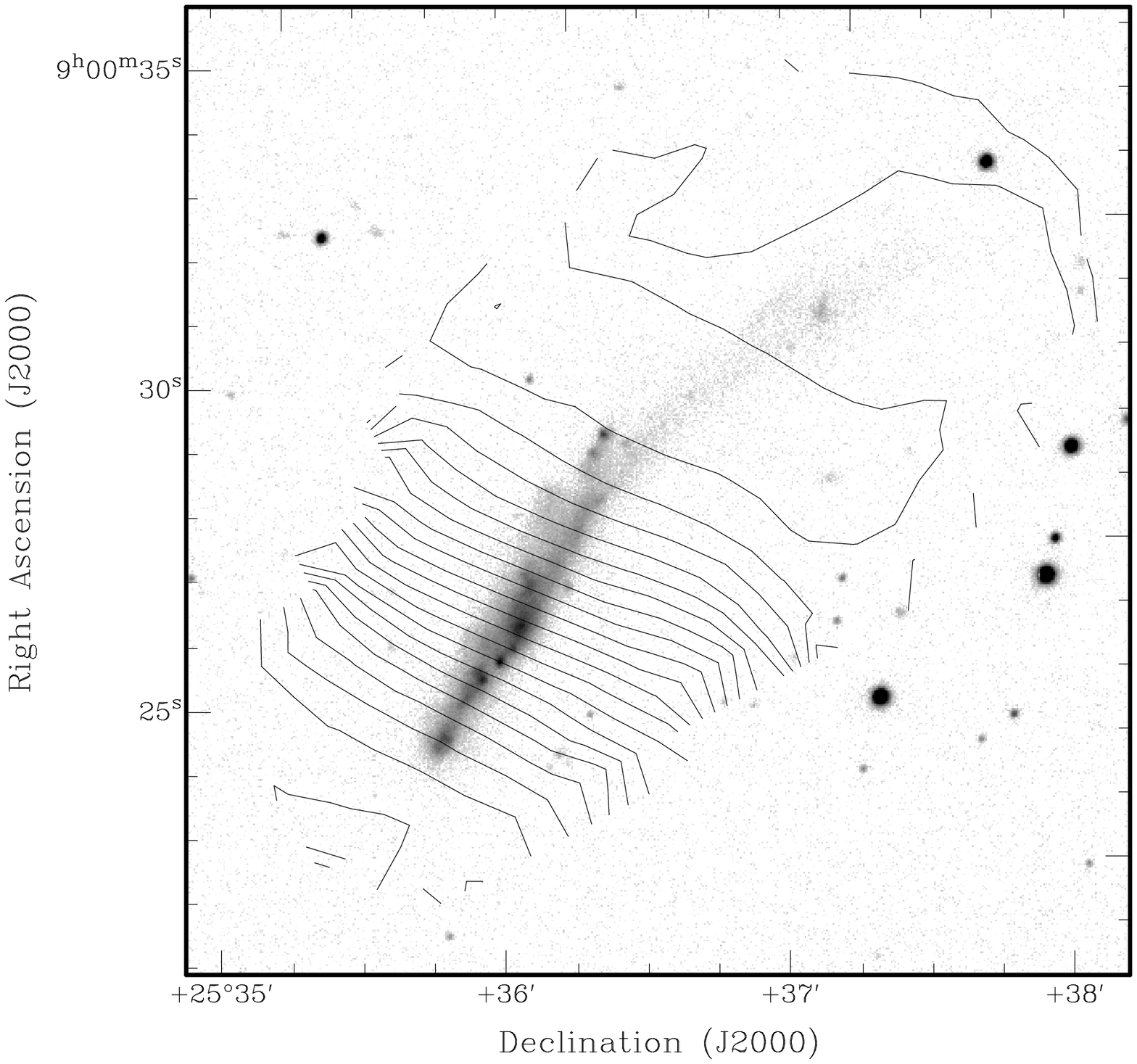}
  \caption{
{\bf Left panel:} Integrated \HI\ emission from the UGC~4722 system 
at a resolution of $\sim$40\arcsec\ (contours) overlaid on the 
SDSS $g$-band image (greyscales). The contours start at 
$2.5 \times 10^{19}$\acc and are spaced a factor of 2 apart.
{\bf Right panel:} similar map, but for the velocity field. The iso
velocity contours start at 1735~\kms and are spaced 5~\kms apart. The
velocity increases smoothly from the SW to the NE.}
	\label{fig:m0+m1.40as}
 \end{figure*}

\begin{figure}
  \centering
\includegraphics[angle=-0,width=6.5cm, clip=]{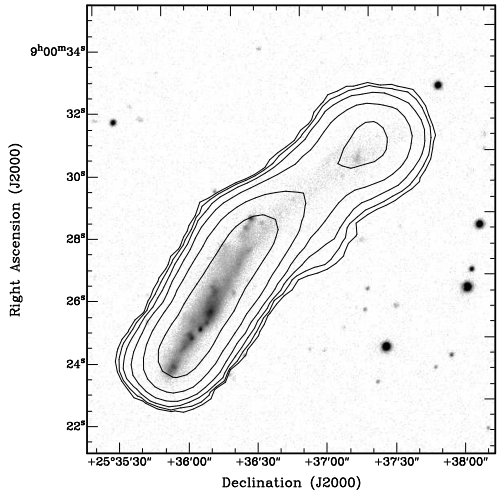}
\caption{
Integrated \HI emission from the UGC~4722 system (contours)
at a resolution of $\sim$25\arcsec\,  overlaid on the SDSS $g$-band image
(greyscales) The \HI\ contours start at $6 \times 10^{19}$~\acc and are spaced
a factor of 2 apart.}
\label{fig:uv10m0ov}
\end{figure}

\begin{figure*}
  \centering
\includegraphics[angle=-0,width=7.5cm, clip=]{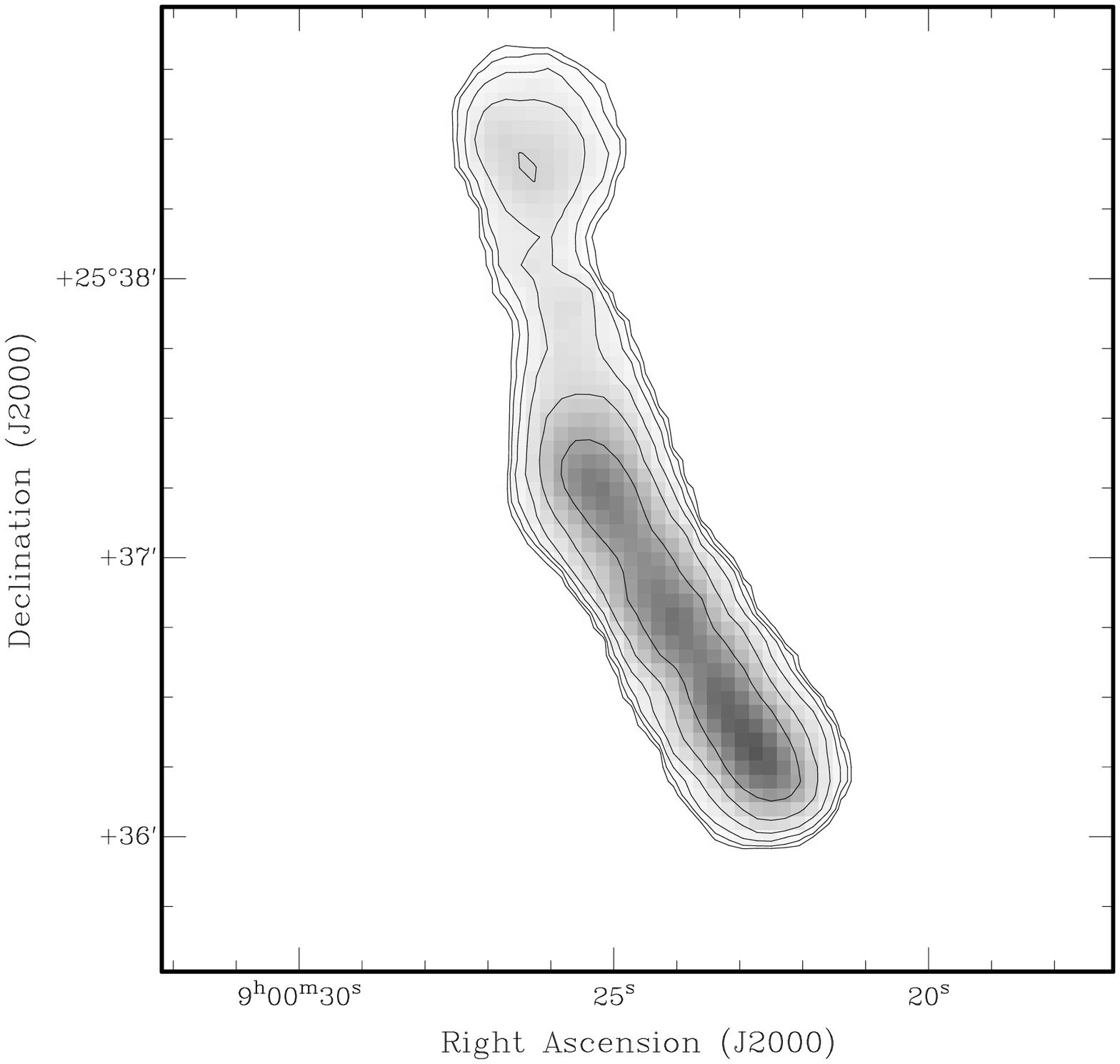} \hskip 0.5in \includegraphics[angle=-0,width=7.5cm, clip=]{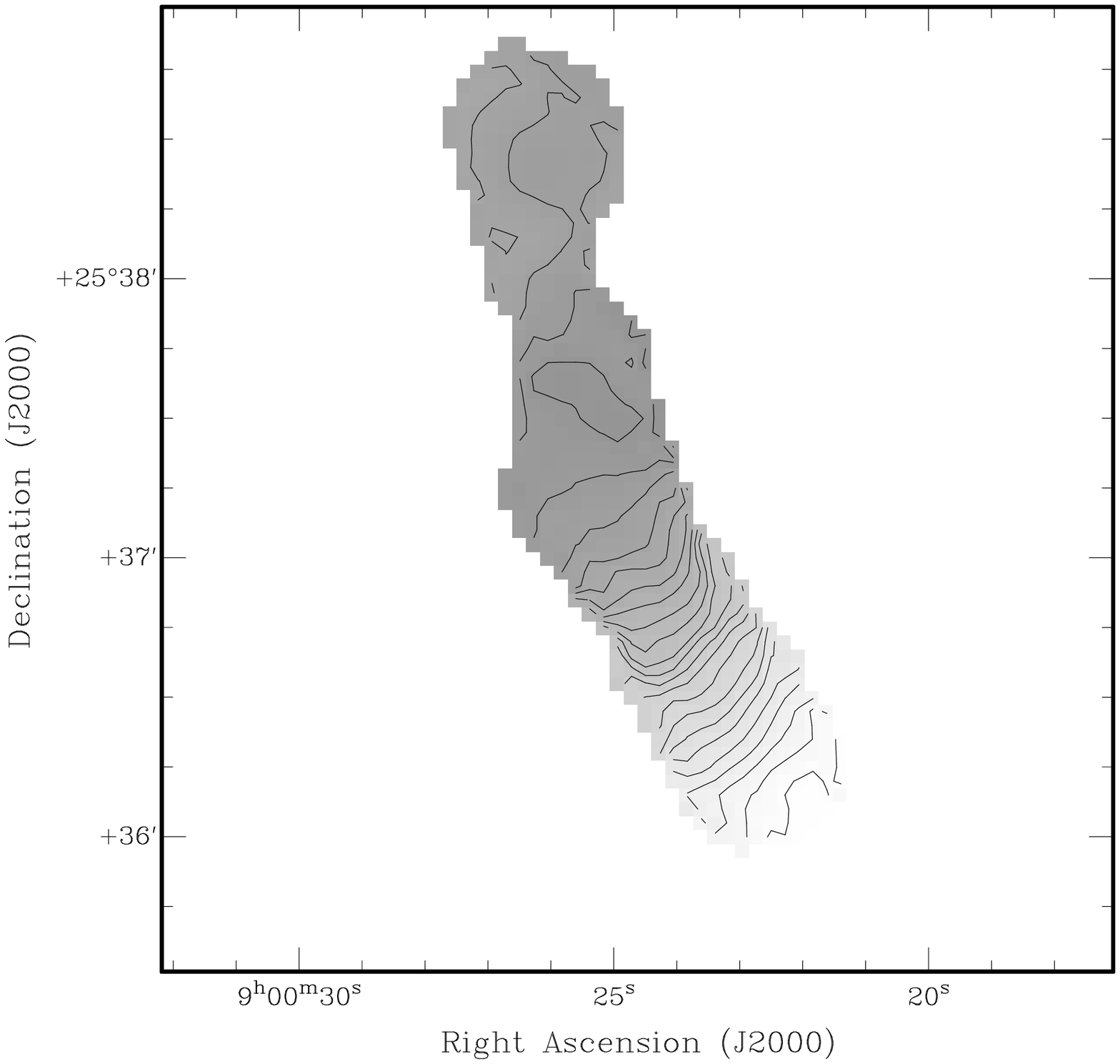}
  \caption{
{\bf Left panel:} Integrated \HI\ emission for a beamwidh of 12\arcsec (both
contours and greyscales). The contour levels start at $1 \times 10^{20}$\acc,
and are spaced apart by a factor of 2.
{\bf Right panel:} \HI\ velocity field at a resolution of 12\arcsec. The
contours start at 1750\kms and are spaced 5\kms apart.}
	\label{fig:m0+m1.12as}
 \end{figure*}

The integrated \HI\ maps at different resolutions ($\sim$40\arcsec, 25\arcsec,
and 12\arcsec) are shown superimposed on the SDSS $g$-filter image in Figs.
\ref{fig:m0+m1.40as},\ref{fig:uv10m0ov} and \ref{fig:m0+m1.12as}. From the 
figures it appears
clear that this system consists of two interacting galaxies, with an
\HI-bridge connecting them. This bridge coincides with
the optical plume  and the \HI\ concentraion corresponds to the nebulosity
`C' at the end of the plume. In the higher resolution
images, one can see that the peak of the \HI\ emission is slightly offset
from the location of the optical nebulosity. The plume (or bridge connecting
 both galaxies) most probably consists of material  pulled out from the
smaller component. We denote this second galaxy as UGC~4722C. The 
\HI-velocity field at resolutions of $\sim$40\arcsec\
and 12\arcsec\ are shown in Figures \ref{fig:m0+m1.40as},\ref{fig:m0+m1.12as}.
At the higher resolution there is a hint that the kinematical major axis 
of \HI-gas near component `C' is perpendicular to the kinematical major 
axis of the main galaxy. We note that another possible interpretations is for 
component `C' to be a tidal dwarf galaxy, with the tidal features being
produced by the gravitational interaction of UGC~4722 with an unseen dark 
galaxy. While this possibility cannot be ruled out without detailed numerical
simulations, a more straight forward explanation is what is proposed
above, i.e. that the system consists of a pair of merging dwarf galaxies. 

\begin{figure}  
  \centering
\includegraphics[angle=-90,width=7.5cm, clip=]{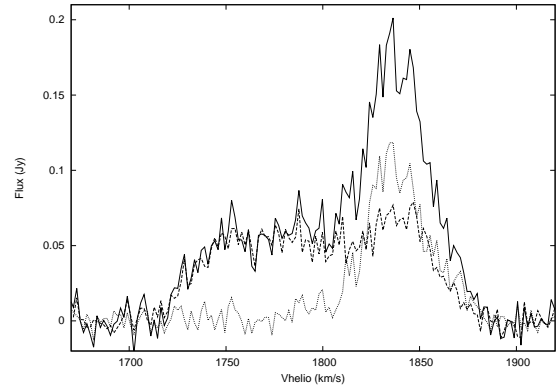}
  \caption{
The integrated \HI-profile of UGC~4722 obtained from
40\arcsec\ resolution datacube (solid line). Also shown are the integrated
spectrum of the region corresponding to the disk of the main galaxy (dashed
line) and the integrated spectrum of the plume (dotted line). 
}
	\label{fig:HI_prof}
 \end{figure}

The integrated \HI-profile (at 40\arcsec\ resolution) of UGC~4722 is shown in
Fig.~\ref{fig:HI_prof}, along with the profile for the main galaxy disk
and the profile of the plume. The total integrated flux measured from
this profile is 12.1$\pm$1.0 Jy~\kms, which is about $\sim$87\%
of the single dish flux measured by the Arecibo Telescope (Haynes et al., 
2011). The quoted error bar includes the uncertainity in the absolute 
flux calibration. From a comparison of the GMRT and Arecibo profiles (not
shown), it appears that some gas near the average velocity of the system
has been resolved out. For the main galaxy disk the integrated \HI\ flux
is 7.6 Jy~\kms, the central velocity is 1795$\pm$3~\kms\ and the velocity 
widths at the 50\% and 20\% levels are  $W_{\mathrm 50}$ = 130$\pm$3~\kms,
$W_{\mathrm 20}$ = 150$\pm$3~\kms. For the plume the corresponding
quantities are 4.3 Jy~\kms, V$_{\rm helio}$ = 1837$\pm$3~\kms and 
$W_{\mathrm 50} \sim$38$\pm$3~\kms.
In case that most of the emission resolved-out in the GMRT observaitons
relates to the main disc, its \HI\ flux can be as large as to 9.6 Jy~\kms. Its
total mass $M$(\HI) and $M$(\HI)/$\L_{\rm B}$ ratio would then be $\sim 25\%$
larger than what is listed in Tab.~\ref{tab:param}.

\subsection{Photometric properties}

\begin{table}
\caption{Photometric parameters of UGC~4722, UGC~4722C and the Plume}
\label{tab:photo}
\begin{tabular}{lccc} \\ \hline \hline
Parameter                           & UGC~4722         & UGC 4722C        &  Plume      \\ \hline
$g_{\rm tot}$                       & 14.70$\pm$0.01   & 19.26$\pm$0.02   &  16.94$\pm$0.01 \\
$(u-g)_{\rm tot}$                   & 0.96$\pm$0.01    & 0.56$\pm$0.05    &  1.08$\pm$0.03  \\
$(g-r)_{\rm tot}$                   & 0.25$\pm$0.01    & 0.11$\pm$0.03    &  0.13$\pm$0.01  \\
$(r-i)_{\rm tot}$                   & 0.10$\pm$0.01    & --0.15$\pm$0.04  &  0.09$\pm$0.02  \\
$(u-g)_{\rm tot,0}$                 & 0.91$\pm$0.01    & 0.51$\pm$0.05    &  1.03$\pm$0.03  \\
$(g-r)_{\rm tot,0}$                 & 0.20$\pm$0.01    & 0.06$\pm$0.03    &  0.08$\pm$0.01  \\
$(r-i)_{\rm tot,0}$                 & 0.08$\pm$0.01    & --0.17$\pm$0.04  &  0.07$\pm$0.02  \\
$B_{\rm tot}$                       & 15.01$\pm$0.02   & 19.53$\pm$0.03   & 17.21$\pm$0.02  \\
$(b/a)_{\rm 25}$                     & 0.22             & 0.50            &  --    \\  \hline
$\mu_{g}$(mag~arcsec$^{-2}$)      & 24.04$\pm$0.28   & 25.20$\pm$0.02   &  24.80           \\
$\mu_{r}$(mag~arcsec$^{-2}$)      & 24.40$\pm$0.15   & 24.99$\pm$0.02   &  24.60           \\
$\mu_{B}$(mag~arcsec$^{-2}$)      & 24.15$\pm$0.32   & 25.50$\pm$0.02   &  25.10           \\
$\mu_{B,c,i}$(mag~arcsec$^{-2}$)  & 24.47$\pm$0.32   & ---              &  ---           \\
$n_{\rm g}$(Sersic)                 & 1.23$\pm$0.24    & ---              &  ---             \\
$\alpha_{\rm g}$($\arcsec$)         & 7.3$\pm$1.8      & ---              &  ---              \\
\hline \hline
\multicolumn{4}{p{8.0cm}}{%
(1) -- values of the total magnitudes are not corrected for the MW
foreground extinction; (2) -- for UGC~4722, $\mu$ corresponds to the
central brightness of the model disc $\mu_{0,g}$ (see Paper IV),
while for UGC~4722C - to the effective surface brightness
$\mu_{eff,g}$- a mean within the effective  radius, and for the plume
to that within a circular aperture with $R$=7.9\arcsec.
 }
\end{tabular}
\end{table}

The photometric parameters measured from the SDSS images are given in
Table~\ref{tab:photo}. The table contains the measured total magnitudes 
$g_{\rm tot}$, colours $(u-g), (g-r)$ and $(r-i)$, the derived $B_{\rm tot}$,
as well as the axial ratio $b/a$ for the both galaxies. Also we show
the total magnitude of the plume emission,  the derived central surface
brightness in $g$ and $r$ filters for UGC~4722, the corresponding central
brightness in $B$, transformed using the formulae given in Lupton et al.
(2006), as well as this parameter corrected for the Milky Way extinction
\citep{Schlafly11} and inclination. Finally, the model fit parameters for
UGC~4722 are included: the Sersic profile index and its characteristic
radius.

For component `C' and the total emission of `plume+C' the surface brightness
relates to model-independent parameters, since it is rather difficult to
apply models to their surface brightness distributions. Thus, for `C' the
$\mu_{\rm eff}$ in filters $g,r,B$ refers to the  mean within the effective
radius, while for the plume this is simply the mean within a circular aperture
with $R =$7.9\arcsec.

\subsection{Oxygen abundances}

\begin{table}
\centering{
\caption{Line intensities in spectra of UGC~4722 and UGC~4722C}
\label{tab:Intens1}
\begin{tabular}{lcccc} \hline \hline
\rule{0pt}{10pt}
& \MC{2}{c}{UGC~4722~`f'} & \MC{2}{c}{UGC~4722C}  \\ \cline{2-3} \cline{4-5}
\rule{0pt}{10pt}
$\lambda_{0}$(\AA) Ion                    &
$F$/$F$(H$\beta$)&$I$/$I$(H$\beta$)       &
$F$/$F$(H$\beta$)&$I$/$I$(H$\beta$) \\ \hline

3727\ [O\ {\sc ii}]\  & 3.75$\pm$0.79 & 2.71$\pm$0.82 & 4.10$\pm$0.30 & 4.15$\pm$0.35 \\
4340\ H$\gamma$\      & 0.14$\pm$0.06 & 0.42$\pm$0.45 & 0.34$\pm$0.03 & 0.47$\pm$0.05 \\
4861\ H$\beta$\       & 1.00$\pm$0.20 & 1.00$\pm$0.37 & 1.00$\pm$0.06 & 1.00$\pm$0.07 \\
4959\ [O\ {\sc iii}]\ & 1.18$\pm$0.22 & 0.86$\pm$0.22 & 0.25$\pm$0.04 & 0.23$\pm$0.04 \\
5007\ [O\ {\sc iii}]\ & 2.69$\pm$0.40 & 1.95$\pm$0.40 & 0.82$\pm$0.07 & 0.73$\pm$0.07 \\
6548\ [N\ {\sc ii}]\  & 0.02$\pm$0.03 & 0.01$\pm$0.03 & 0.05$\pm$0.03 & 0.04$\pm$0.02 \\
6563\ H$\alpha$\      & 3.61$\pm$0.51 & 2.78$\pm$0.61 & 3.45$\pm$0.16 & 2.78$\pm$0.16 \\
6584\ [N\ {\sc ii}]\  & 0.09$\pm$0.05 & 0.07$\pm$0.05 & 0.20$\pm$0.04 & 0.16$\pm$0.03 \\
6716\ [S\ {\sc ii}]\  & 0.46$\pm$0.09 & 0.33$\pm$0.09 & 0.51$\pm$0.05 & 0.40$\pm$0.04 \\
6730\ [S\ {\sc ii}]\  & 0.32$\pm$0.08 & 0.23$\pm$0.08 & 0.36$\pm$0.05 & 0.28$\pm$0.04 \\
& &  &  & \\
C(H$\beta$)\ dex     & \MC {2}{c}{0.00$\pm$0.18} & \MC {2}{c}{0.16$\pm$0.06}  \\
EW(abs)\ \AA\        & \MC {2}{c}{3.50$\pm$2.25} & \MC {2}{c}{3.7$\pm$0.31}   \\
$F$(H$\beta$)$^a$\   & \MC {2}{c}{0.65$\pm$0.09} & \MC {2}{c}{2.29$\pm$0.09}          \\
EW(H$\beta$)\ \AA\   & \MC {2}{c}{9.12$\pm$1.26} & \MC {2}{c}{33.88$\pm$1.55} \\
V$_{hel}$\ \kms\     & \MC {2}{c}{1860$\pm$42}   & \MC {2}{c}{1830$\pm$54}    \\ \hline \hline
\MC{5}{l}{$^a$ in units of 10$^{-16}$ ergs\ s$^{-1}$cm$^{-2}$.}
\end{tabular}
 }
\end{table}

\begin{table}
\centering{
\caption{Derived physical parameters and Oxygen abundances}
\label{t:Chem1}
\begin{tabular}{lcc} \hline \hline
\rule{0pt}{10pt}
\rule{0pt}{10pt}
Value                                 & UGC~4722~`f'      & UGC~4722C        \\ \hline
%
$T_{\rm e}$(OIII)(10$^{3}$~K)\        & 16.11$\pm$1.86   & 16.56$\pm$1.13   \\
$T_{\rm e}$(OII)(10$^{3}$~K)\         & 14.39$\pm$1.98   & 14.54$\pm$1.19   \\
$N_{\rm e}$(SII)(cm$^{-3}$)\          &   10$\pm$10~~    & 10$\pm$10~~     \\
&   & \\
O$^{+}$/H$^{+}$($\times$10$^{-4}$)\   & 0.280$\pm$0.149  & 0.414$\pm$0.113  \\
O$^{++}$/H$^{+}$($\times$10$^{-4}$)\  & 0.193$\pm$0.062  & 0.062$\pm$0.011  \\
O/H($\times$10$^{-4}$)\               & 0.474$\pm$0.162  & 0.476$\pm$0.113  \\
12+log(O/H)\  (IT07)                  & 7.68~$\pm$0.15~  & ~7.68$\pm$0.10~  \\
12+log(O/H)\  (PM2010)                & 7.59~$\pm$0.21~  & ~7.42$\pm$0.10~  \\
12+log(O/H)\  (weigh. mean)           & 7.64~$\pm$0.06~  & ~7.50$\pm$0.10~  \\
\hline \hline
\end{tabular}
 }
\end{table}

\begin{figure}
\centering
\includegraphics[angle=-90,width=8cm, clip=]{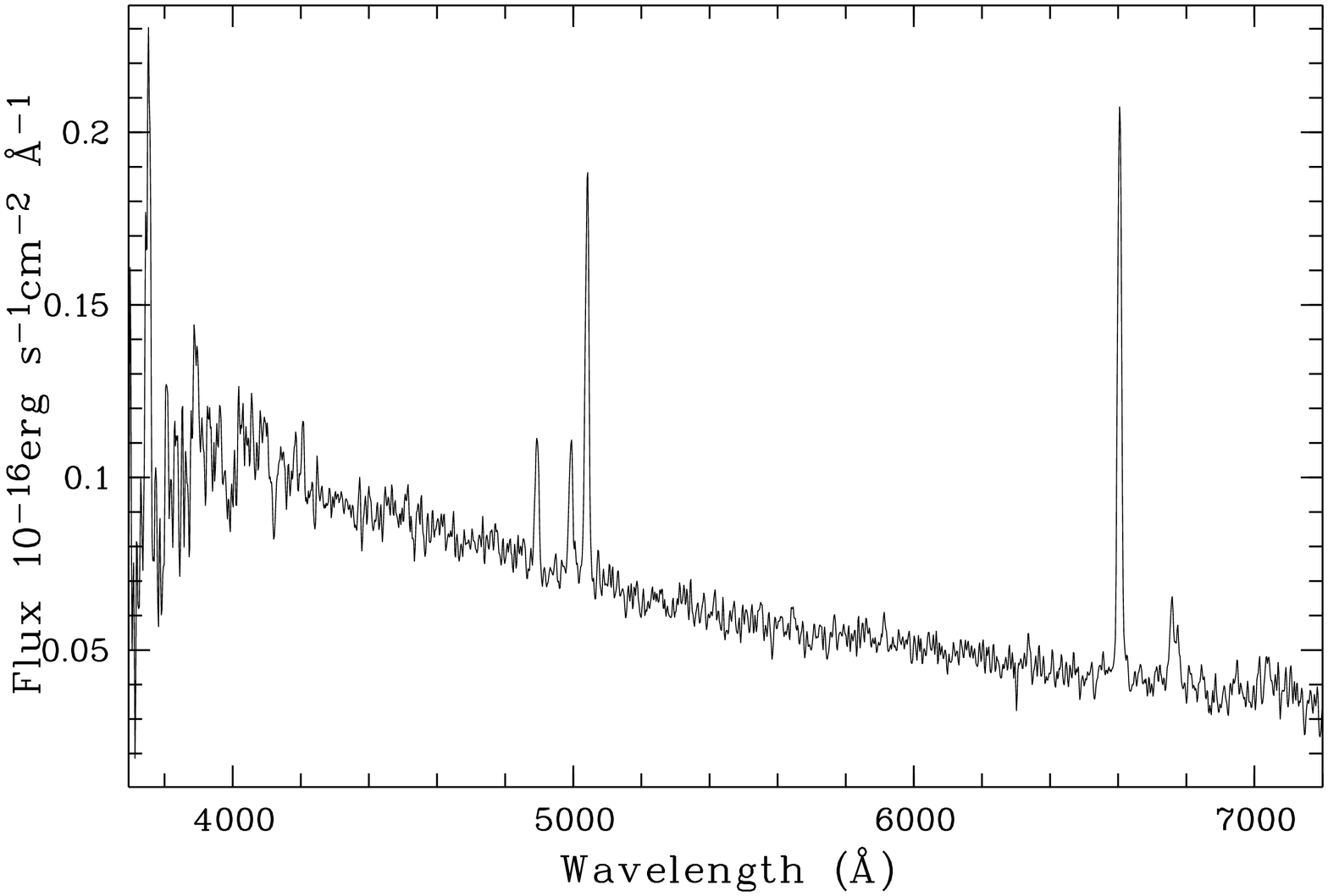}
\includegraphics[angle=-90,width=8cm, clip=]{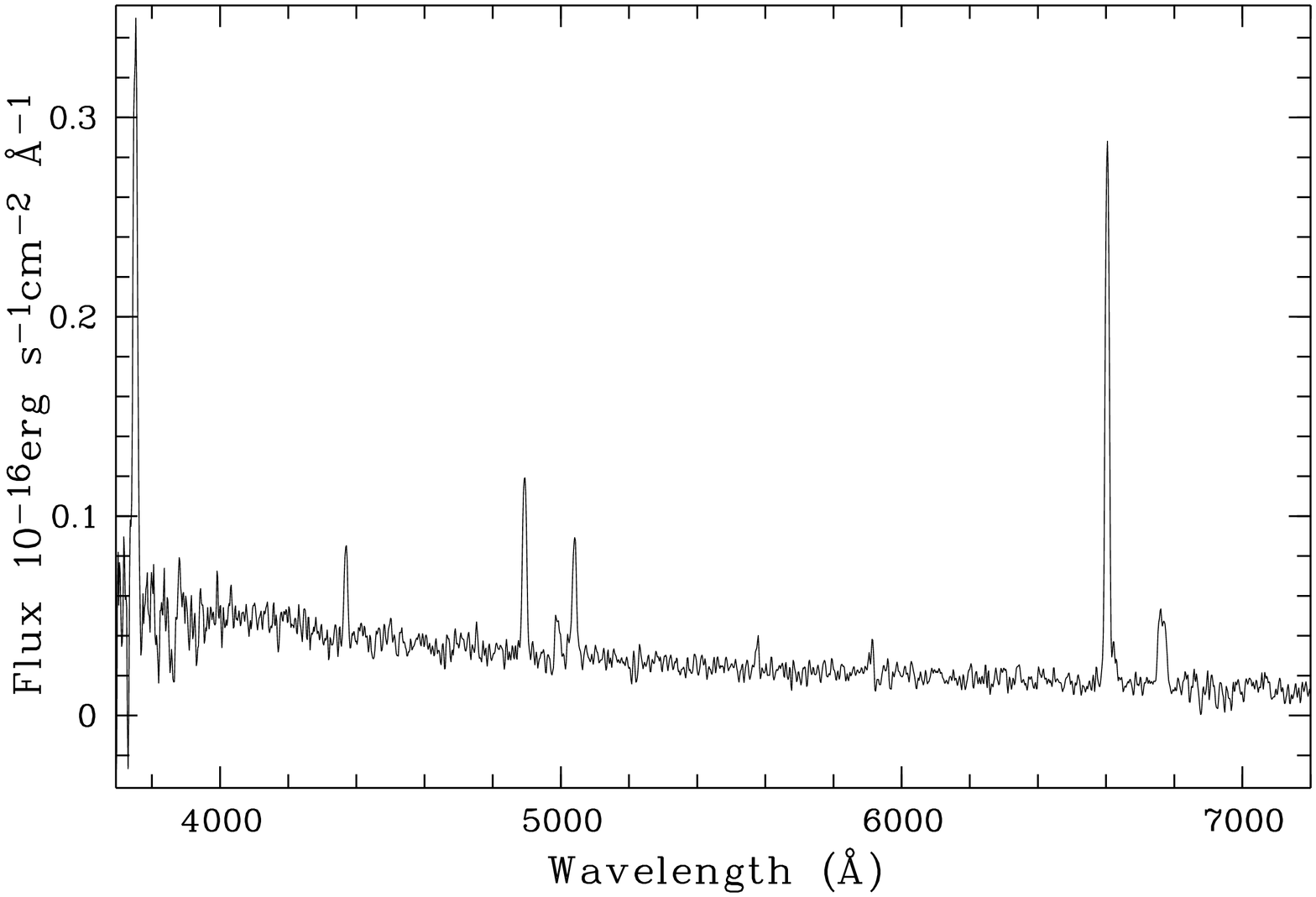}
\caption{
The BTA mid-resolution spectra of UGC~4722 system (see regions in Fig.~\ref{fig:image}).
Top panel: 1D-spectrum of the bright HII-region on the northern edge of UGC~4722 (region `f').
Bottom panel: 1D-spectrum of the nebulosity `C'.
}
\label{fig:spectra}
\end{figure}

After processing the long-slit spectrum, described in Sec.~\ref{sec:obs}
({\bf No.~1,} grism VPHG550G), we obtained a 2D spectrum, from which we
extracted
for further analysis two 1D spectra with the largest signal-to-noise
ratio for the strong emission lines. This was done using standard techniques 
\citep[see e.g.,][for a detailed description]{Kniazev08}. In
Fig.~\ref{fig:spectra}~(top) we show the extracted 1D spectrum for the 
brightest central part (extent $\sim 1.8\arcsec$) of the prominent \HII-region 
in UGC~4722 (region `f'), while in Fig.~\ref{fig:spectra}~(bottom) we show 
the extracted 1D spectrum for the brightest central part of the faint 
nebulosity `C' (extent $\sim 3.2\arcsec$). The measured line intensities, used 
for determination of the oxygen abundance O/H are listed in the upper part of 
Table~\ref{tab:Intens1}. In the bottom of the Table we also give the derived
parameters of the extinction coefficient C(H$\beta$), the equivalent width of 
the Balmer absorption (EW(abs)), the total flux in the H$\beta$ line, the
equivalent width of the H$\beta$ emission (EW(H$\beta$))and the weighted value
of the radial velocity, derived from redshifts of strong emission lines.

Both the spectra are rather noisy, so it was possible to measure
fluxes only for the strongest emission lines. In particular, given the 
modest signal-to-noise and equivalent widths  of [OIII]$\lambda$4959
line, it is understandable that we do not detect the temperature-sensitive 
[OIII]$\lambda$4363 line, which is normally a factor of 12 to 30 times
fainter in the $T_{\rm e}$ range of 20000~K to 13000~K.
We are hence unable to use the classical $T_{\rm e}$ method to estimate
O/H from  these spectra. Instead we used the well-calibrated and reliable 
semi-empirical method of \citet{IT07} as well as the empirical estimator 
suggested by \citet{PM11}, both of which use only the intensities of 
the strong lines.

The resulting physical parameters and the derived values of
O/H with their errors are given in Table~\ref{t:Chem1}.  The parameters
listed include
the electron temperature in the main O$^{++}$ emission zone,
$T_{\rm e}$(OIII) (derived via the semi-empirical method), the temperature 
in the O$^{+}$ zone, $T_{\rm e}$(OII), and the adopted value of the 
electron density $N_{\rm e}$. The derived relative abundances of O$^{+}$, 
O$^{++}$ and O along with their  errors are shown in the next three entries.
Finally, we show the parameter 12+$\log$(O/H) derived via semi-empirical 
method as well as with empirical O/H estimator of \citet{PM11}. We also 
calculated the weighted mean of these two values, which we use for 
all further analysis below.

\section[]{Discussion}
\label{sec:dis}

As discussed above, UGC~4722 is located in the nearby Lynx-Cancer void and
is one of the most isolated galaxies known. It is a member of both the 
Catalogue of Isolated Galaxies (CIG) as well as the AMIGA sample
of isolated galaxies. Our \HI\ and optical data however suggest that this
system actually consists of a pair of interacting dwarf galaxies.  This
indicates that the selection criteria for these samples of isolated
galaxies allow  at least some minor merger type systems to be classified 
as isolated galaxies. The finding that UGC~4722 consists of a galaxy pair,
as well as the other discovery of pairs and small groups of dwarf galaxies
in nearby voids is also consistent with the suggestion of \citet{Kreckel12} 
that the small scale clustering in voids is similar to that in the general
field \cite{KM08,MK09}. Below we discuss in the detailed properties of 
the UGC~4722 system.

\subsection{Main parameters}

In Table~\ref{tab:param} we present the main parameters of the UGC~4722 system. 
From the total magnitudes in the $g$ and $r$ filters (Table \ref{tab:photo}) and 
the transformation equations between SDSS magnitudes and other systems 
by Lupton\footnote{\url{https://www.sdss3.org/dr8/algorithms/sdssUBVRITransform.php\#Lupton2005}},
we derived the total $B$-band magnitudes, $B_{\rm tot}$=15.01$\pm$0.02 and
19.53$\pm$0.03 for the main galaxy and the nebulosity `C' respectively.
As mentioned above, we assume that the plume consists of material pulled out
from the fainter galaxy. It is then natural to add the integrated light of
the plume to that of the nebulosity `C' itself, in order to get a robust
estimate of the total luminosity of the companion. These total magnitudes, 
designated as C+plume, are also listed.

Following \citet{PaperI} in assuming that this system has a large peculiar
velocity  (as suggested earlier by \citet{Tully08}), we adopt its distance modulus
of $\mu$=32.22 ($D$=27.8 Mpc). We note that the distance adopted here is
close to the average value of 26.2~Mpc listed in NED (which is derived
from the Tully-Fisher relation). The corresponding absolute magnitudes are
hence M$_{\rm B}^0 = -$17.38 (main galaxy),  $-12.86$ (nebulosity `C')
and --15.18 (`C' + plume). This implies that the ratio of blue luminosities
of the two interacting galaxies is $\sim$7.7 (in contrast to the ratio of
$M$(\HI)$\sim$1.8). The $M$(\HI)/$L_{\rm B}$ ratio (in solar units) for
the galaxies is $\sim$1.0 for the main galaxy and $\sim$4.3 for `C'+plume.
The estimate of the minimal dynamical mass of the main disc UGC 4722, adopting
its distance, \HI-radius of 50\arcsec\ and $V_{\rm rot}$=75~\kms\ (i.e.
half of the 20\% width W$_{20}$) gives the value of $M_{\rm dyn}$(UGC~4722) 
$\sim 8.5 \times 10^{9}$ $M$\sunn. For component `C' both because of the
highly disturbed nature of the velocity field, as well as the fragmented
nature of the object we are unable to estimate the dynamical mass.

\subsection{Plume properties}

To better understand the properties of the plume, we examine the  SDSS-based
surface brightness, magnitudes, colours and their gradients along the plume. 
The $ugri$ colours do not show visible gradients along the plume. Therefore 
it is reasonable to work with the integrated plume parameters. The integrated
$ugri$ colours, (after accounting for the Galaxy foreground extinction 
\citep{Schlafly11}) are as follows: \mbox{$(u-g)_{0}$} = 1.03$\pm$0.03,
\mbox{$(g-r)_{0}$} = 0.08$\pm$0.01, \mbox{$(r-i)_{0}$} = 0.07$\pm$0.02.

\begin{figure*}
  \centering
\includegraphics[angle=-90,width=16.0cm, clip=]{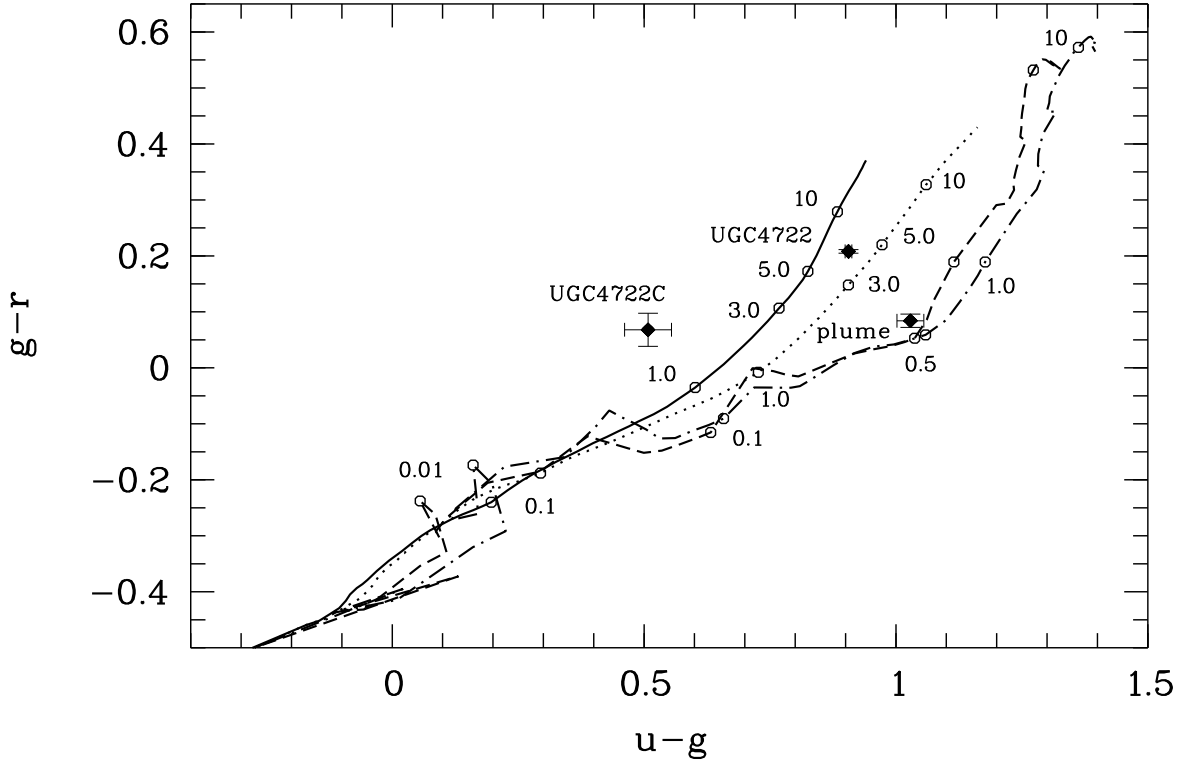}
  \caption{
PEGASE2 evolutionary tracks for constant SFR and 
instantaneous SF and metalliicty of $z$=0.002. 
Solid line corresponds constant SFR with Salpeter (1955) IMF;
dot -- constant SFR with Kroupa et al. (1993) IMF;
dash -- instantaneous SF with Salpeter IMF;
dot-dash -- instantaneous SF with Kroupa IMF.
Octagons along the tracks (with corresponding numbers) mark the time elapsed since the beginning of SF (in Gyr). 
The integrated colours corrected for Galactic extinction are shown by black diamonds with error bars and 
are marked by object names (UGC~4722 -- the main Sdm galaxy, 
UGC~4722C -- compact emission-line nebulosity at the plume edge, 
plume -- total emission of the plume and nebulosity `C'). 
See Sec.~\ref{sec:dis} for details. 
We note that the colours  of nebulosity `C' are affected by the nebular emission related to its resent and current SF activity, 
however it has negligible effect on the integrated colours of the plume.
}
\label{fig:ugr}
\end{figure*}

As can be seen in Fig.~\ref{fig:ugr}, these colours are an excellent match
to the \mbox{PEGASE2} \citep{pegase2} evolutionary model tracks for instantaneous SF
(or simple stellar population, SSP) for both Salpeter (1955) and Kroupa
et al. (1993) Initial Mass Functions (IMF) and metallicity $z$=0.002 (roughly
$Z=Z$\sunn/10). The latter $Z$ is close to the metallicity found in the
central \HII-region of the nebulosity `C'. The tracks for both types of
IMF run very
close to each other at this time range (around 0.5~Gyr). Our best fit
estimate of
the time since the starburst corresponds to  $\sim$0.45-0.5~Gyr with a 
Salpeter IMF. Observational studies show that the peak in star
formation activity in companion galaxies occurs near the pericenter 
passage (e.g. \citet{patton13}. Numerical models (e.g., GalMer, 
\citet{dimatteo08}) also make similar predictions, albeit for
specific initial conditions.
Our estimated star burst time hence corresponds approximately
to the time since the last close encounter between the two galaxies.
This in turn allows us to estimate the relative velocity of the companion in
the sky plane. Dividing the projected distance between the centres of 
UGC~4722 and the nebulosity `C' (i.e. $\sim$13.8~kpc) by the time
since perigalactic passage (i.e. $\sim$500~Myr), one obtaines 
$V_{\rm trans} \sim$28~\kms. This is about 0.6 of the difference in radial 
velocities between the two components.

From the PEGASE2 models one can derive many parameters related to the
stellar population including the temporal dependence of the specific 
luminosity (per solar mass) in several passbands. Adopting these parameters
at an epoch of T=500~Myr, as well as the solar absolute magnitude of
$M_{\rm V}$=3.78 and the colour index
$g-V$=--0.11, we get $M_{\rm g}$=3.67.  For the estimated 
plume absolute magnitude $M_{\rm g,plume}$= --15.38, one then derives a
total stellar mass of 4.6$\times$10$^{7}$~$M$\sunn. This can be
compared to the total \HI-mass ($M$(\HI)= 78$\times$10$^{7}~M$\sunn)
or a total gas mass ($M$(\HI+He)=1.33$\times$$M$(\HI)),
$M_{\rm gas}=104\times10^{7}~M$\sunn. That is the plume's visible stellar
mass comprises only $\sim$4\% of its baryonic mass.


\subsection{Gas Phase metallicities}

We obtained two  O/H measurements in this system: one in the central
\HII-region of the nebulosity `C' and the second - in the bright
\HII-region on the edge  of the UGC~4722 disk (region `f'). Both the measured
values are rather similar, with the formal difference ($\sim$0.08 dex) being
comparable to the uncertainties ($\sim$0.07 and 0.12 dex). We note
that differences in O/H of as large as $\sim$0.25--30~dex (i.e. up to 
factor of two) would be consistent with our error bars. Significantly 
higher S/N-ratio spectra of these \HII\ regions would be needed to
rule out this possibility.

We also checked whether these measured metallicities are consistent with
the luminosity-metallicity (``L-Z'') relation derived for galaxies residing 
in denser environments. In particular, we compare their metallicities with
that expected from the relation derived for the Local Volume dwarfs with
well-known distances and O/H by \citet{Berg12}. The scatter in this relation
is quite small, $\sigma_{\rm (O/H)}$=0.15~dex. The comparison shows that the 
measured O/H in our two dwarfs are lower than what would be expected from
the L-Z relation by a factor of $\sim$3 for
the main galaxy UGC~4722 and a factor of $\sim$2 for the
smaller companion UGC~4722C. The deviation is significantly larger than
the measurement errors. While this deviation may partly result from
gas infall from the `unevolved' outer parts of the galaxies due to
the strong tidal disturbance \citep[as suggested by][]{EC10,montuori},
we also note that galaxies in the Lynx-Cancer void in general
have a lower value (by $\sim$ 30--50\%) of O/H, \citep{void_OH} than 
galaxies residing in denser regions.

\subsection{Origin of UGC~4722 and comparison with other similar systems}

The suggestion in \cite{ikar08} 
that the UGC4722 system represents the interaction between a dwarf galaxy and
a ``dark'' galaxy now appears unlikely, since the baryonic mass in the plume
($\sim 10^8$M$_\odot$) is larger than the mass of the `dark' halos expected 
in the LCDM models. Instead, a plausible scenario for 
the UGC4722C system is that it is the result of a minor merger between 
UGC4722 and an extremely gas rich companion which has come in on a parabolic
orbit. Had the companion been on a bound orbit, one would have expected that
the extended tidal interaction could have led either to a disruption of the
galaxy or that tidally induced star formation would have led to a larger 
stellar mass fraction. After the closest approach of the parabolic orbit 
$\sim$0.5 Gyr ago, some of the gas ($\sim$4\% ) collapsed and transformed 
to stars while a significant fraction of the gas was also pulled out to 
form the ``plume''. In this scenario it is interesting to try and determine
what the properties of the smaller companion galaxy were before the start
of the collision.  From the currently available optical data, it is difficult
to estimate the mass of the stellar population of the companion with ages 
older than 1~Gyr. If there were to be such a population, one would expect that
it could also be pulled out in the plume but would remain undetectable
because of its very low surface brightness. It seems reasonable to assume
however that any such population would have a stellar mass similar to that
of the stellar mass of the  `old' stellar population related to UGC~4722C,
which is $\sim$7 times less luminous than the plume. This implies that 
before the collision the companion was an extremely gas-rich dwarf. In the 
most extreme case, when the most of its stars formed in the course of the 
current collision, the pre-collision stellar mass fraction could be as low 
as 1\% or smaller. While this is indeed unusual, it would not be a unique
case, as several similar very gas-rich dwarfs have already found in this
void \citep{void_LSBD}.  In fact the Lynx-Cancer void hosts two even more 
extreme gas-rich dwarfs \citep{CP2013} . An alternative model to the minor
merger is that the UGC4722 represents a very clumpy gas disk. While the
disks of somewhat brighter gas rich galaxies are known to be somewhat
clumpy  \citep[e.g.][]{Wang2013}), cases as disturbed looking as UGC4722
are rare. Additionally, the kinematics of the UGC4722 system are clearly
discrepant from what one would expect from a clumpy disk.

The only similar morphology galaxy which we found in the literature is
DDO~169 (UGC~8331). This is a nearby dwarf irregular galaxy ($M_{\rm B} 
\sim -$13.8, $D_{\rm TRGB}$=4.4~Mpc), which has been used in many statistical
studies. The only detailed paper dedicated to this galaxy by \citet{IS06} 
is based on rather shallow \HI\ VLA observations (10 and 20 minutes of
integration
in C and D configurations) and H$\alpha$ imaging. The earlier \HI-maps 
obtained with Westerbork Radio Telescope also provide only limited 
sensitivity\footnote{\url{http://www.nrao.edu/astrores/HIrogues/}}.
\citet{IS06} did not measure  the colours of the plume, or those of the 
brighter knots on the plume edge, and detected H$\alpha$ emission only near
the bright central body of the galaxy. In the VLA maps one can see that
there is a \HI-peak that coincides with an optical knot at the end
of the plume. UGC~8331 and UGC~4722 hence appear fairly similar, but
a more detailed study of UGC~8331 will be needed to establish this in
detail.
There are also two other objects with tidal tails/plumes that resemble
the UGC~4722 system. One of the best known is the Tadpole galaxy
(UGC~10214=VV~29), studied in detail by \citet{Briggs01}. Several very LSB
plumes near large disc galaxies are found by \citet{MBD11} via deep
analysis of the SDSS images. Some of them can be analogs of UGC~4722 case.

Although the number of such objects known appears limited, we
note that the UGC~4722 system has many simiarlities with other small
dwarf aggregates found in voids. Like UGC~4722, many void pair and
triplet
members, show elevated values of $M$(\HI)/$L_{\rm B}$ ($\sim$2 to 25), 
blue colours of their outer regions and low O/H. Examples include: the merger
remnant/triplet DDO~68 \citep{Ekta08,Tikhonov14,Cannon14}, 
the triplet J0723+36 \citep{CP2013}, the pair HS~0822+3542/SAO~0822+3545 
\citep{HS0822}, pair J0852+13 and the triplet MRK~407 (Chengalur et al., 
in prep.). A more detailed comparison of these unusual systems in voids 
is postponed to the separate paper. We also note that there are very 
few detailed numerical simulations of the merger of extremely gas rich 
small systems. Such studies would be an important theoretical counterpart
to the observations of UGC~4722 and the other systems discussed above. 
Understanding the merger of small gas rich objects is also important 
to a more detailed understanding of herirachical galaxy formation models.

\begin{table}
\caption{Main parameters of UGC~4722 and UGC~4722C}
\label{tab:param}
\begin{tabular}{lcc} \\ \hline
Parameter                          & UGC 4722               &  UGC~4722C            \\
R.A.(J2000.0)                      & 09 00 23.54            & 09 00 26.11           \\
DEC.(J2000.0)                      & +25 36 40.6            & +25 38 21.4           \\
$A_{\rm B}$ (from NED)             & 0.17                   & 0.17                  \\
$B_{\rm tot}$                      & 15.01$^{(1)}$          & 19.53$^{(1)}$         \\
$B_{\rm tot}$                      &                        & 17.21 (C+Plume)       \\
$V_{\rm hel}$(HI)(\kms)            &1795$\pm$1$^{(1)}$      & 1837$\pm$4$^{(1)}$    \\
$V_{\rm hel}$(opt)(\kms)           &1780$\pm$5$^{(1)}$      & 1795$\pm$5$^{(1)}$    \\
$V_{\rm LG}$(opt)(\kms)            &1714                    & 1729                  \\
Distance (Mpc)                     &27.8$^{(1)}$            & 27.8$^{(1)}$          \\
$M_{\rm B}^0$                      & --17.38$^{(1)}$        & --12.86$^{(1)}$       \\
$M_{\rm B}^0$                      &                        & --15.18 (C+Plume)     \\
Opt. size (\arcsec)$^{(5)}$        & 90$\times$20$^{(1)}$   & 10$\times$5$^{(1)}$   \\
Opt. size (kpc)                    & 12.1$\times$2.7        & 1.35$\times$0.67      \\
$\mu_{\rm B,c,i}^0$(mag~arcsec$^{-2}$) & 23.35  $^{(1)}$    & ---                   \\
12+$\log$(O/H)                     & 7.64$\pm$0.06$^{(1)}$  & 7.50$\pm$0.10$^{(1)}$ \\
\HI\ int.flux$^{(6)}$              & 7.6$\pm$1.2$^{(2)}$    & 4.3$\pm$0.2$^{(2)}$   \\
$W_\mathrm{50}$ (km s$^{-1}$)      & 130$\pm$3$^{(2)}$      & 38$\pm$3$^{(2)}$      \\
$V_\mathrm{rot}$ (\kms)            & 75$^{(3)}$             & 43$^{(4)}$            \\
$M$(\HI) (10$^{7} M$\sunn)         & 138$^{(1)}$            & 78$^{(1)}$            \\
$M_{\rm dyn}$ (10$^{8} M$\sunn)    & 85$^{(1)}$           & $-$           \\
$M$(\HI)/$L_{\rm B}^{(7)}$         & 1.0$^{(1)}$            & 4.3$^{(1)}$           \\
\hline
\multicolumn{3}{p{8.0cm}}{(1) -- derived in this paper;
(2) -- derived from the GMRT \HI-profile. Real \HI-flux and mass of UGC~4722
can be
$\sim$25\% larger; (3) -- derived from the GMRT \HI-profile and (4) --
from H$\alpha$ P--V diagram (UGC~4722C); (5) -- $a \times b$ at
$\mu_{\rm B}=$25\fm0~arcsec$^{-2}$; (6) -- in units of Jy$\cdot$\kms; (7) --
in solar units. 
}
\end{tabular}
\end{table}

\section{Summary and Conclusions}

\begin{enumerate}
\item Analysis of GMRT \HI\ images indicates that UGC~4722 is a pair of merging
  dwarfs, as opposed to the interaction of a dwarf galaxy with a ``dark'' 
  companion. The main galaxy has a rotational velocity 
  $V_{\rm rot} \sim$ 75~\kms,
  The smaller companion galaxy is almost completely disrupted making its 
  rotation velocity hard to measure, our best estimate from the \HI\
  data is $V_{\rm rot} \sim$35~\kms.
\item Analysis of H$\alpha$ $P-V$ diagrams obtained with the slit positioned
  along the optical plume and the compact emission-line nebulosity on its edge,
  supports the conclusion
  that the nebulosity is an individual dwarf galaxy with a rotation velocity
  $V_{\rm rot} \sim$40~\kms.
\item The combination of the optical photometry data, obtained from the SDSS
  images, with the estimates of the total \HI\ fluxes derived from the GMRT
  \HI-maps, reveals that both components are gas-rich, with 
  $M$(\HI)/$L_{\rm B}$ of 1.0 and 4.3, for UGC~4722 and the nebulosity `C' 
  plus plume, respectively.
\item A comparison of the plume colours with the PEGASE2 model evolutionary
  tracks for instantaneous SFR shows that the colours are well matched by 
  a stellar population produced in a single burst about $\sim 0.5$~Gyr ago. 
  As suggested by statistical and numerical studies of starbursts in 
  interacting galaxies we assign the time of the starburst to the 
  peri-galactic passage of the companion.
\item The oxygen abundance in the \HII-regions in the two galaxies are
  12+$\log$(O/H)=7.64$\pm$0.06 (UGC~4722) and 7.50$\pm$0.10 (UGC~4722C). 
  For both UGC~4722 and UGC~4722C the measured metallicity is substantially 
  (i.e. a factor of 2--3) lower than that expected from the L--Z relation
  for similar luminosity galaxies in denser environments.

\end{enumerate}

\section*{Acknowledgements}

The authors thank A.V.~Burenkov for help with BTA observations.
SAP, DIM, YAP and ESS acknowledge the partial support of this work through
RFBR grant No.~14-02-00520.
DIM, IDK and SAP acknowledge partial support from RFBR grant No.~13-02-90407.
JNC, IDK, SAP and DIM acknowledge the RFBR-Indian collaborative grant
No.~13-02-92690.  We acknowledge useful comments from the referee,
N.~Trentham, which have helped improve this paper. 
The authors acknowledge the photometric data and the related
information available in the SDSS database used for this study.
This research has made use of the NASA/IPAC Extragalactic
Database (NED), which is operated by the Jet Propulsion Laboratory,
California Institute of Technology, under contract with the National
Aeronautics and Space Administration. This paper is based in part on 
observations taken with the GMRT. We thank the staff of the GMRT who 
made these observations possible. The GMRT is run by the National Centre 
for Radio Astrophysics of the Tata Institute of Fundamental Research


\bsp

\label{lastpage}

\end{document}